# Structures and Superconductivity of Hydrogen and Hydrides under Extreme Pressure

Zihan Zhang[1], Wendi Zhao[1], Defang Duan[1, *], Tian Cui[1, 2, *]

[1]*State Key Laboratory of Superhard Materials and Key Laboratory of Material Simulation Methods & Software of Ministry of Education, College of Physics, Jilin University, Changchun 130012, China*

[2]*Institute of High Pressure Physics, School of Physical Science and Technology, Ningbo University, Ningbo, 315211, People's Republic of China*

**Abstract:**

Metallic hydrogen, existing in remarkably extreme environments, was predicted to exhibit long-sought room-temperature superconductivity. Although the superconductivity of metallic hydrogen has not been confirmed experimentally, superconductivity of hydrogen in hydrides was recently discovered with remarkably high critical temperature as theoretically predicted. In recent years, theoretical simulations have become a new paradigm for material science, especially exploration of material at extreme pressure. As the typical high-pressure material, metallic hydrogen has been providing a fertile playground for advanced simulations for long time. Simulations not only provide the substitute of experiments for hydrogen at high-pressure, but also encouraged the discovery of almost all the experimentally discovered superconducting hydrides with the record high superconducting transition temperature. This work reviews recent progress in hydrogen and hydrides under extreme pressure, focusing on phase diagram, structures and the long-sought goal of high-temperature superconductivity. In the end, we highlight structural features of hydrides for realization of hydrogen-driven superconducting hydrides near ambient pressure.

# Contents

# 1. Introduction

Hydrogen, the simplest element in universe, is destined to be one of the most important elements, playing a role in exploration nature. For example, the wave function and energy spectrum of a single hydrogen atom could be analytically solved from Schrödinger equation, which has opened the door for people to quantum mechanics a hundred years ago. Meanwhile, hydrogen has always been one of the most important testing systems for various models and numerical simulation methods of quantum solid theoretical physics[1-4]. For example, the pressure-driven metallization of hydrogen was predicted shortly after the development of quantum solid-state theoretical physics in 1935[1], and the potential of high-temperature superconductivity of metallic hydrogen was identified shortly after construct of Bardeen-Cooper-Schrieffer theory in 1968[5]. Although the high-temperature superconductivity of metallic hydrogen has not been confirmed, the idea of hydrogen-driven superconductivity directly leads to the discovery of hydride superconductors, which break the critical temperature records of cuprate superconductors. Compared with other superconductors, the research course of hydrogen-based superconductivity is different, where theoretical simulations play the unprecedented role.

Although hydrogen is the simplest element, the phase diagram of hydrogen is remarkably complex. As shown in Fig. 1 (a), pressure and temperature are two thermodynamic pathways to metallic hydrogen. During cold compression of hydrogen, hydrogen becomes metallic and exhibits long-sought room temperature superconductivity and high-energy density. Therefore, metallic hydrogen is hailed as the long dreamed of "holy grail" in the high-pressure physics. Besides high pressure, temperature also triggers the dissociation of molecular hydrogen and leads to metallization through liquid-liquid phase transition. Hydrogen in universe generally exist in the high-temperature environment such as outer mantle of Jupiter. The research for behaviors of temperature-driven atomic hydrogen is crucial for current models of giant planets[6,7]. Moreover, in the research on inertial fusion, the equation of state of hydrogen needs to be clarified through numerical simulation before compression into the region of deuterium-tritium fusion[8].

Metallic hydrogen provides a possible route to high temperature superconductivity, but the high stable pressure limits further experiments and applications. Therefore, "chemical pre-compression" theory was proposed for moderate-pressure superconductivity via "pre-compressed" atomic hydrogen in hydrides[9], and the next challenge is how to search hydride superconductors from millions of hydrides. Theoretical simulations provide a possible route to accurately predict the structures and properties of potential high-temperature superconductors and successfully guide high-pressure experimental synthesis[10]. For example, covalent hydride $H_3S$ and clathrate hydrides were synthesized shortly after their theoretical predictions[11-15]. Since "chemical pre-compression" theory was proposed 20 years ago, hydride superconductors had achieved great success and exerted a powerful influence on the related research field, such as high-pressure experimental technique[16], superconducting theory[17] and material design[10]. However, stable pressure of atomic hydrogen sublattices in hydrides still limits their further research and application of hydrogen-driven superconductivity. The next challenging goal in the research field of superconducting hydrides is design of structural templates of superconducting hydrides, which exhibit both high-temperature superconductivity and structural stability at ambient pressure as shown in Fig. 1 (b).

Aimed at realization of high-temperature superconductivity of hydrides at ambient pressure, theoretical exploration for hydride superconductors has been carried out including developments of "chemical pre-compression" theory[18] and high-throughput calculations[19-22], which screen potential hydride superconductors from millions of structures from thousands of hydrogen-rich systems. As the results of theoretical developments, structural features of hydrides, which keep a good balance between superconductivity and stability, were identified to guide high-pressure experiments. For example, "hydrogen-based alloy backbone" theory[18] was proposed to structural design for hydride superconductors at moderate pressure by engineering binary X-H backbones, which directly led to the synthesis of first ternary hydride superconductor $LaBeH_8$ with a resolved crystal structure[23]. The high-throughput calculations for hydrides were carried based on structures from typical hydride family and structure search[19], including clathrate hydrides[20], perovskite structure[22] and fluorite-type backbone[24]. Moreover, with the deepening exploration of hydride superconductors, this relationship of chemical compression theories and high-throughput calculations is more and more intensive, which are promotion and complementarity for each other. Many hydrides were discovered with high-temperature superconductivity at ambient pressure in recent high-throughput calculations structures, such as $Mg_2IrH_6$[19,21] and $Na_2AuH_4$[22]. Notably, the recently reported structures of superconducting hydrides, which keep a good balance between superconductivity and stability, generally consist of "pre-compressing" cations and "pre-compressed" H-rich backbones as description of "hydrogen-based alloy backbone" theory. Therefore, high-throughput calculations driven by chemical compression theories would play an important role in the future exploration for hydride superconductors, especially at moderate pressures.

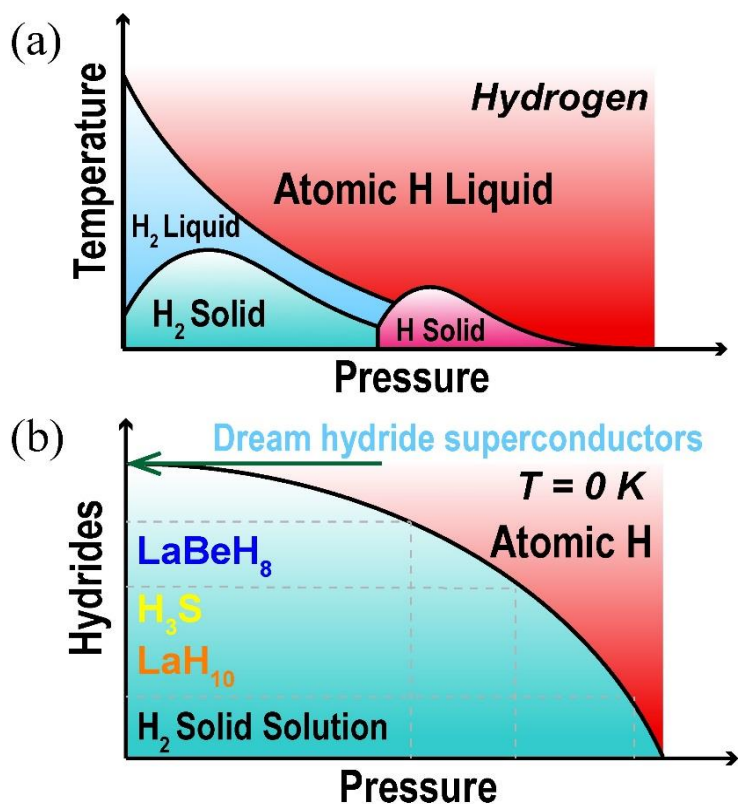

Fig. 1 Sketches of phase diagram of hydrogen (a) and explorations of structural template of hydride superconductors (b).

In this review, we summarize the research achievements of exploration for structure and superconductivity of hydrogen and hydrides under high pressure, and the important role of theory and computation therein. From the achievements of design of hydrides with hydrogen-driven superconductivity, structural features of hydride superconductors were summarized for further exploration of high-temperature superconductivity at ambient pressure. We start by the exploration for structure and superconductivity of hydrogen at high pressure in Sec. II. We then summarized the achievements of superconducting hydrides at high pressure in Sec. III. Finally, we discuss the empirical rules from the exploration of the structure and superconductivity of hydrogen and hydrides, and predict the tendency for future development of hydride superconductors in Sec. IV.

# 2. Structures and Potential Superconductivity of Hydrogen at High Pressure

Hydrogen is the most simple and abundant element in the universe, and attracted wide interests from condensed matter physics. Metallic hydrogen has traditionally been regarded as the Holy Grail of high-pressure science for the potential room-temperature superconductivity of metallic hydrogen. However, exploration of high-pressure structures of hydrogen remains challenging for the limitation of high-pressure experiments and strong quantum effect of crystal lattice in theoretical simulations. This section reviews the achievements of research on structures and potential superconductivity of hydrogen at high pressure

## 2.1 The research of hydrogen at high pressure

The prediction of metal hydrogen under high pressure in 1935 was the first non-empirical prediction for material solely based on quantum mechanics and physical constants[1]. In the early 20th century, quantum solid-state theoretical physics was not yet the mainstream research direction in the field of theoretical physics. However, in the early 1930s, it was an era of vigorous development in quantum solid theory and physics, providing prerequisites for non-empirical prediction of materials based on quantum mechanics.In 1927, the one of the most important approximations in solid physics, Born-Oppenheimer approximation, was proposed, which separated the motion of atomic nuclei and electrons to simplify the many-body systems in condensed matter physics. In 1928, Felix Bloch proposed the Bloch wave function, which describes the behavior of wave functions in periodic lattices. In 1929, John C. Slater proposed the Slater polynomial, which describes the expression of the wave function in multi electron systems. Then the Hartree-Fock equation was proposed for multi-electron systems by Vladimir A. Fock. In 1930, Léon Brillouin proposed the Brillouin zone, which further provided a prerequisite for the numerical calculation of electron wave functions in lattice of crystals. Based on these theory, Wigner's research group in Princeton university had tried the first non-empirical prediction of materials. After proposing the Wigner Seitz primitive cell, Wigner and his doctoral student Frederick Seitz calculated the electronic structure and mechanical properties of solid sodium and lithium[25-27]. Although the results of calculation are rough limited by the numerical computational ability at the time, the Wigner team still achieved electronic structure calculations based on quantum mechanics, as well as predictions from electronic structure to material properties.

In 1931, physicist Alan Wilson proposed the rules to distinguish between conductors and insulators based on whether energy bands pass through Fermi surface, and proposed the Wilson phase transition from insulators to metals driven by high pressure. Subsequently, physicist J. D. Bernal believed that all materials could become metallic under high pressure. Based on Wilson and Bernal's theory and computational experience of the electronic structure and mechanical properties for solid sodium and lithium[25-27], Wigner and his doctoral student Hillard B. Huntington predicted the metallization of hydrogen. The electronic structure and mechanical properties of the atomic hydrogen phase were calculated with body centered cubic (bcc) structure. The bcc phase of metallic hydrogen was predicted to be stable at high pressure ~ 250000 atmospheres[1]. From the current research perspective, this pressure is far lower than the actual pressure for molecular hydrogen to enter metallization. But there is no doubt that prediction of metallic hydrogen is an important attempt to non-empirically simulate material based on quantum solid theory physics, providing the possible route to predict properties of materials from electronic structure.

With the development of computer hardware, computational physics has emerged as one of the important disciplines in physics[28], and hydrogen has played an indispensable role in the development of computational physics[29]. Because of the simple electronic structure of hydrogen, many approximations in solid-state physics simulations can be omitted. For example, freeze of core electrons with the pseudopotential approximation is not necessary, so the error brought by the pseudopotential is small in the calculation for electronic structure of hydrogen. And the electronic structure can even be directly calculated using the real Coulomb potential $1/r$ without the pseudopotential approximation. Moreover, the orbital angular momentum of hydrogen is very weak, so there are almost no relativistic effects such as spin-orbit coupling. The Schrödinger equation can well describe the electronic structure of hydrogen. Therefore, hydrogen is an important testing playground for solid physics numerical simulations based on density functional theory or quantum Monte Carlo[29,30].

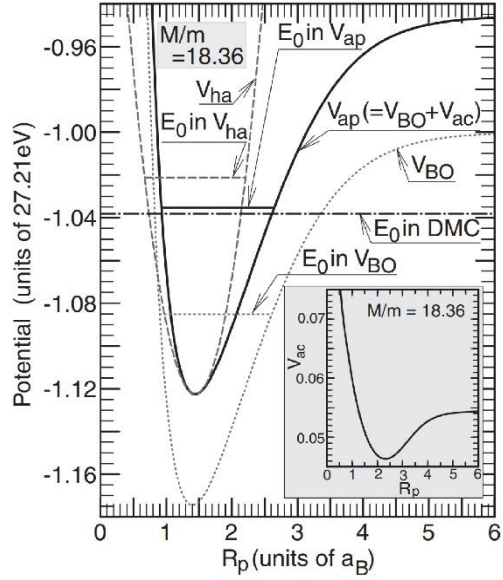

Fig. 2 Four-body diffusion Monte Carlo simulation of hydrogen molecules with a 100-fold increase in electron mass[30]. $V_{BO}$ is the Bonn-Oppenheimer potential energy surface. $V_{ac}$ is the adiabatic correction of kinetic energy operator between the electrons and protons ($V_{ac}(r_p) = \langle \varphi_e | \hat{T}_p | \varphi_e \rangle_e$), which cannot be ignored after the increase of electron mass. $V_{ap}$ is the effective potential for protons, which can be represented by the sum of $V_{BO}$ and $V_{ac}$. $V_{ah}$ is harmonic potential under the harmonic approximation.

Although the electronic structure of hydrogen is simple, simulation for hydrogen is still difficult because the Born-Oppenheimer approximation might be not applicable to hydrogen. Because the nucleus of hydrogen is a proton with mass 1836 times heavier than electrons, nucleus of hydrogen is also a particle that exhibits quantum effects. The description of the nuclear quantum effect of hydrogen in the Born-Oppenheimer approximation is difficult, mainly reflected in two aspects: (i) It may not be reasonable to directly separate wave functions between atomic nuclei and electrons, and the impact of the strong coupling needs to be considered. (ii) The zero point kinetic energy of atomic nuclei is extremely high (quantum effects make the kinetic energy of quantum particles non-zero, and under harmonic approximation, the kinetic energy is $E_k = \frac{1}{2}\hbar\omega$). The Bonn-Oppenheimer potential energy surface completely ignores the kinetic energy of the atomic nucleus. In response to this issue, Professor Yasutami Takada in the University of Tokyo and Professor Tian Cui carried out diffusion Monte Carlo simulations of hydrogen molecules under four-body Coulomb interactions[30], where the mass ratio of protons and electrons is variable, in order to evaluate the validity of the Born-Oppenheimer approximation, as shown in Fig. 2, where electron mass of electron was increased by 100 times in this simulation. The error caused by the first aspect can be represented by the difference between the ground state energy of diffusion Monte Carlo ($E_0$ in DMC) and the energy level of the proton in the effective potential ($E_0$ in $V_{ap}$). It can be seen that even if the electron mass is increased by 100 times, the difference between the two energy levels is still small. The error caused by the second aspect can be represented by comparing the energy level of the proton in the effective potential ($E_0$ in $V_{ap}$ in the figure) with the minimum value of the effective potential. The difference in this energy level does not depend on the mass of the electron, but is only caused by the low mass of the proton itself.

For the first excess of Born-Oppenheimer approximation in simulation of hydrogen, diffusion Monte Carlo simulation with the four-body Coulomb interaction of hydrogen molecules reveals that when the mass ratio of electrons to protons is less than 1:10, the error caused by separation of protons and electrons does not exceed 1% in calculated the ground state energy of hydrogen molecules. It is suggested that the motion of atomic nuclei and electrons can be separated to calculate the ground state energy in the case of molecular hydrogen. However, the electronic density of molecular hydrogen is localized in the midpoint of hydrogen-hydrogen bonds in real space, and pressure causes electrons to tend to delocalize. The delocalized electrons exhibit huge screening ability against nuclear vibrations. Therefore, there is still no clear conclusion on the evaluation of the error caused by the separation of delocalized electrons and the movement of atomic nuclei. For the second excess of Born-Oppenheimer approximation in simulation of hydrogen, it indicates that the kinetic energy of hydrogen cannot be ignored, and it still maintains a huge kinetic energy (about 1000 K) even at low temperatures. Therefore, when hydrogen condenses into a solid, the phase diagram of solid hydrogen always exhibits complex, including both the molecular rotation and intermolecular interactions of hydrogen molecules before dissociation under high pressure[3], and the proton correlation before atomic hydrogen after molecular dissociation, all exhibit rich dynamic configurations[31].

## 2.2 The pressure-temperature phase diagram of hydrogen

The scale of temperature and pressure in condensed matter physics is much lower than that in high-energy physics and plasma physics, so the temperature and pressure range of metal hydrogen phase transition is not considered extreme (below 10000 K and 1 TPa) compared with high-energy physics. It is reported that there are two pathways for metallization of hydrogen: the transition from an insulating molecular hydrogen solid to a metal atomic hydrogen solid, and from an insulating molecular hydrogen liquid to a metal atomic hydrogen liquid. In this section we discuss these two pathways of transformation.

There are mainly five solid phases before the atomic metal phase of hydrogen, as shown in Fig. 4. Phase I is the solid phase with the lowest pressure, where hydrogen molecules are arranged in a densely packed hexagonal (hcp) lattice with random orientations at hcp lattice positions[32]. As pressure increasing and temperature decreasing structure of hydrogen will transform to phase II from phase I[32]. The characteristic of phase II is that the coupling between molecules is enhanced and the free rotation of molecules on the lattice is restricted, as the results of pressure increasing. Therefore, phase II of hydrogen is called the broken-symmetry phase (BSP). Because the phase transition boundary of phase II is sensitive to isotopic effects[32], it can be predicted that the transition of hydrogen from phase I to phase II were predicted to be dominated by nuclear quantum effects in addition to pressure and temperature. The concept of quantum phase transition is well demonstrated in *ab initio* path-integral molecular dynamics simulation carried out for phase II[3].

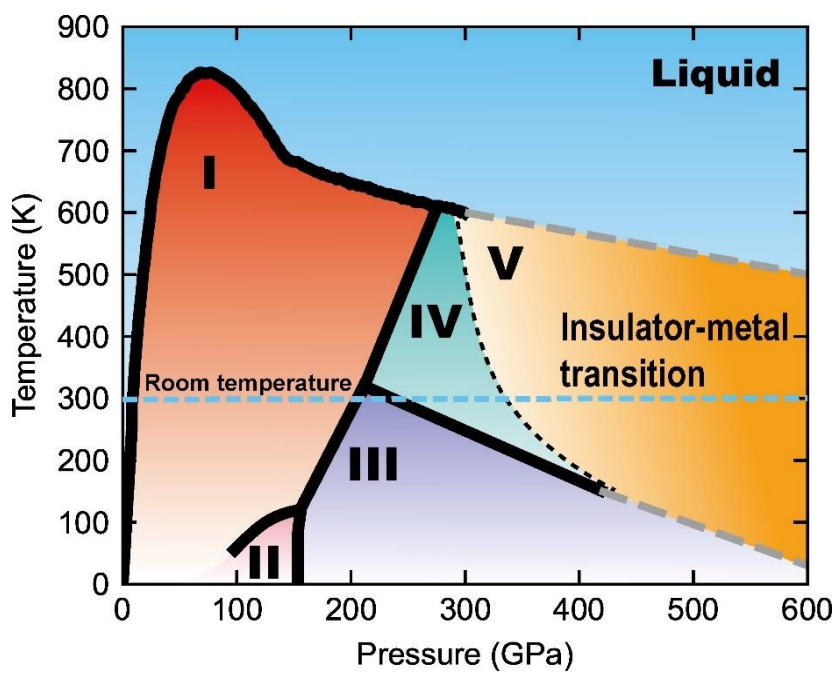

FIG 3 Temperature-pressure phase diagram of solid hydrogen and the main high-pressure phases of hydrogen.

In 1989, the phase III of hydrogen was discovered[33]. In 2011 and 2012, Phase IV was discovered by two different laboratories [34,35]. Compared with the transition from phase I to II, the transition of hydrogen into phases III and IV is relatively stable, with weak isotopic effects. Therefore, these two types of phase transitions were predicted to be dominated by the Born-Oppenheimer potential energy surface. Therefore, the structure search[36-38] plays an important role in identification for the structures of phases III and IV[4,39]. The main difference between structure search and path-integral Monte Carlo or molecular dynamics algorithms is the considerations of nuclear quantum effects for hydrogen. The path-integral Monte Carlo and molecular dynamics algorithms deal with potential energy surfaces that contain nuclear quantum effects, while the structure search randomly generates and optimizes structures on the Born-Oppenheimer potential energy surface, and then the zero energy was considered through harmonic approximation for nuclear quantum effects. In the process of structure search, crystal structures of hydrogen generally were optimized based on Born-Oppenheimer energy surface. Although the Stochastic Self Consistent Harmonic Approximation (SSCHA) provide a possible way to optimize the structures of hydrogen with nuclear quantum effects [40-43]. Due to its huge burden of large computation, it is difficult to carried out a structure search for hydrogen based on SSCHA to comprehensively optimize random structures with the quantum energy surface[44]. Therefore, in identification of the structures of phase III and IV of hydrogen, which was dominated by the Born-Oppenheimer energy surfaces, structure search has played a unique advantage. In 2007, shortly after the structural search was proposed, the candidate structure with a $C2/c$ space group closest to III discovered in 1989 was theorical predicted using *ab initio* random structural search (AIRSS) [4]. In the same year as the discovery of phase IV of hydrogen, candidate structures were soon identified for phase IV of hydrogen using *ab initio* random structural search algorithms[39].

In 2016, Eugene et al discovered phase V of hydrogen in experiment [45], but identification of its crystal structure remains challenging. On the one hand, the experimental measurement is performed at room temperature [45-47] as shown in FIG 3, suggesting that phase V is high temperature phase with proton diffusion driven by quantum effect . In 2013, the dynamic behavior of phase IV with quantum effects was studied using path-integral molecular dynamics[48], and it is suggested that phase IV underwent quantum-effect-driven proton diffusion at room temperature. This phenomenon may indicate that phase V has a diffusing lattice of hydrogen at higher pressure. Because the zero-point kinetic energy of hydrogen might be higher than the diffusion barrier. On the other hand, the electronic structure closed to metallization make phase V a dynamic phase[49,50]. Considering the electronic structure of phase IV, there is a theoretical evidence that phase IV exhibits Dirac semimetallic properties[51]. Despite the low density of electronic states at the Fermi level, electrons have already shown delocalization. The correlation between two hydrogen nuclei intra molecule is screened by delocalized electrons, while the correlation between hydrogen atoms inter molecules has strengthened due to the shortened distance between molecules. Pressure drives electrons to become more delocalized and enhances their ability to screen nucleus vibrations, which is reflected in the fact that the melting point of solid hydrogen decreases with increasing pressure before entering the metal phase[50,52-54]. This phenomenon of decreasing melting point is common in alkali metals[43, 44]. In the pressure range of decreasing melting point of alkali metals, there are a abundant of phases and complex electronic structures[55,56]. The phase V of hydrogen is also in the range where the melting point decreases with pressure. Considering that hydrogen's nuclear quantum effect is much higher than other alkali metals, the difficulty of identification for the structure of phase V may be even higher than that of alkali metals in the decreasing melting point range. The phase V is likely to be the last molecular phase for hydrogen to enter the metal phase, even if no phase transition emerges from phase V to metallic, band gap of phase V will be closed driven by pressure according to Wilson's theory of phase transition.

The structure of metallic hydrogen phase with lowest stable pressure was theoretical predicted with space group of $I4_1/amd$[57,58], which is a candidate structure with the lowest enthalpy value after considering the zero-point energy through harmonic approximation below 1 TPa. The coordination number of hydrogen in $I4_1/amd$ phase is 4, not 8 in bcc metallic hydrogen predicted by Wagner in 1935[1]. The classical metal lattice of body centered cubic and face centered cubic have energy advantages only above 5 TPa [57]. It is suggested that the structure of metallic hydrogen with the lowest stable pressure predicted by structural search is a phase with polymeric hydrogen molecular, similar to the experimentally obtained polymeric nitrogen[59] and polymeric carbon monoxide[60]. However, the valence electrons of polymeric hydrogen are much lower than those of polymeric nitrogen and carbon monoxide, it is difficult to occupy the multicenter bonds of polymeric hydrogen. Because of half-occupied bonds, the metallicity of polymeric hydrogen is stronger than that of polymeric nitrogen and carbon monoxide, but intercepted polymeric hydrogen at low pressure may also become more difficult. Currently, recent numerical simulations based on diffusion Monte Carlo and SSCHA predict a pressure of 577 GPa for hydrogen to enter the $I4_1/amd$ phase[43].

Compared to the difficulty of pressure-induced metallic hydrogen at room temperature, liquid metallic hydrogen has already been obtained through high temperature at different pressure points as shown in Fig 4. The metallization of liquid hydrogen at high temperatures is characterized by different experimental methods, such as high-temperature and high-pressure electrical transport measurement at 80 GPa [61], and ultrafast spectroscopy at 150 GPa [62]. Although temperature can decompose hydrogen molecules and lead to metallization of hydrogen atomic liquids, this temperature-driven insulator-to-metal phase transition still remains a problem. The decomposition of hydrogen molecules seems to be a first-order phase transition with the number of hydrogen molecules as the order parameter[29,49]. However, recent numerical simulations have shown that hydrogen molecules smoothly decompose and produce in the liquid metal phase of hydrogen, achieving dynamic equilibrium. Therefore, the transition from the molecular phase to the atomic phase of liquid hydrogen is a smooth phase transition[63].

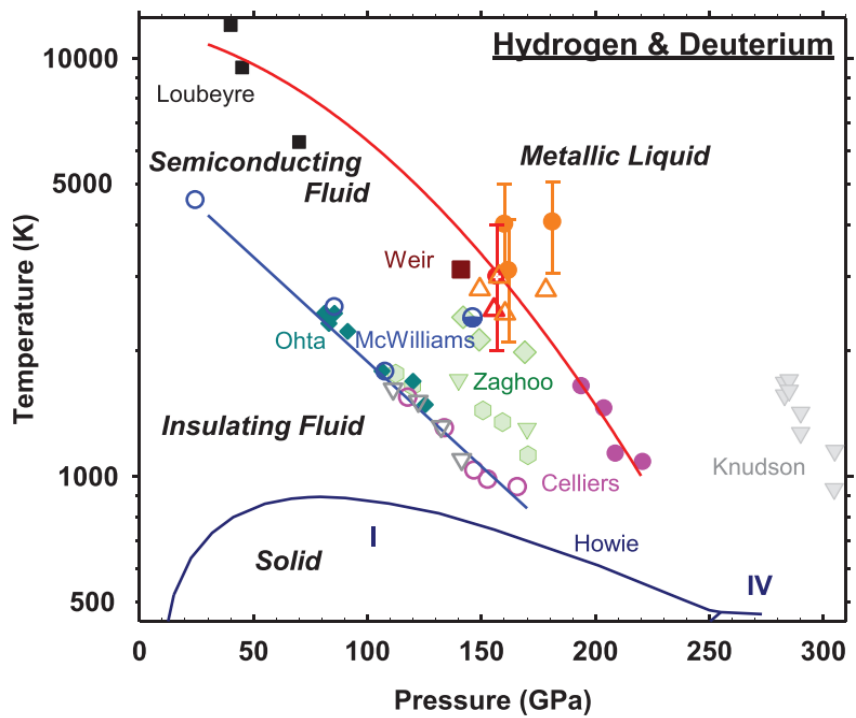

Fig 4 The temperature and pressure phase diagram of hydrogen focused on the temperature induced insulator-to-metal transition of hydrogen.

In exploration of the transition from hydrogen molecular phase to atomic phase, scientists define order parameters in numerical simulations mainly based on the bond length of hydrogen bonds. A representative model[64] is shown as following:

$$\xi(P) = 1 - \frac{R_{H_2-H_2}(P) - r_{H-H}(P)}{R_{H_2-H_2}(P_{1\,atm}) - r_{H-H}(P_{1\,atm})}. \quad (1)$$

Where $r_{H-H}(P)$ and $R_{H_2-H_2}(P)$ are the bond lengths of hydrogen molecules in the system and the shortest distance between hydrogen molecules, respectively. When $\xi(P) = 1$, it indicates that there are no hydrogen molecules in the system and it is completely atomized. When $\xi(P) = 0$, it indicates that the bonding behavior in the system is the same as complete molecule at normal pressure. However, when similar order parameters are applied to the simulation of high-temperature induced phase transitions[63], it is found that the hydrogen-hydrogen bond order parameter does not disappear near the temperature and pressure range of the metallic phase transition, but monotonically decreases, indicating that the temperature induced metallic phase transition of hydrogen is still difficult to describe by order parameters.

## 2.2 Potential superconductivity of metallic hydrogen

After the electron-phonon coupling mechanism was widely believed as a microscopic mechanism of conventional superconductivity, in 1968 Neil W. Ashcroft proposed that metallic hydrogen under high pressure may be a potential high-temperature superconductor [5]. The main analysis process is as following, a formula based on BCS theory can be given:

$$T_c = 0.85\theta_D e^{[-1/N(0)V]}, \quad (2)$$

where $\theta_D$ is the Debye temperature, and $N(0)V$ represents the ability of electron-phonon coupling. Due to hydrogen being the lightest element, the Debye temperature is very high.

However, not all light metals are high-temperature superconductors. At atmospheric pressure the superconducting critical temperatures of Li and Be, two light metals, are 0.0004 and 0.026 K, respectively. On the contrary, the superconducting transition temperature ($T_c$) of the heavy element with a relative atomic mass of 175 reaches 7 K. Therefore, the $N(0)V$ term reflecting the electron-phonon coupling in formula (2) should also be considered. There are two reasons to intuitively estimate the electron-phonon coupling: the scattering ability of the nuclei and the effective charge number of the nucleus. The former can be roughly judged by the real radius of the cation, and the latter can be directly estimated by the valence of the compound. From these two points, it can be inferred that alkali metal atoms have a large radius, but only one valence electron, so $N(0)V$ tends to approach 0. The effective charge number of the nucleus of Be is 2, so the expected $N(0)V$ is larger than that of Li, resulting in a higher expected $T_c$ than that of Li. Other alkali metals and alkali earth metals with larger atomic radii and an effective charge of 1 or 2 have not been found to exhibit superconductivity in current low-temperature experiments. In the third main group of metals (with an effective charge of nuclei is +3), aluminum and gallium also have the lowest $T_c$ of 1 K. The $T_c$ of lead and tin in the fourth main group are higher than those in the third main group. The germanium in the fourth main group and metal elements in the fifth main group exhibit semimetallic properties, which is not conducive to superconductivity. From another perspective, the electrical resistivity at room temperature could generally reflect to the scattering ability of phonons towards electrons. Copper, silver, gold, and platinum with low electrical resistivity at room temperature have not observed superconductivity in current low-temperature experiments, and vanadium, niobium, and tantalum with superconductivity are not the best conductors. Based on nuclear scattering ability and valence electron, although metal hydrogen only has one valence electron, the radius of the hydrogen atom is the smallest. Hydrogen does not have core electrons, and all electrons participate in the scattering of atomic nuclei. The screening ability of valence electrons against nuclear vibrations is extremely strong, so the strength of electron-phonon coupling of atomic metallic hydrogen may be unprecedented, surpassing all currently known metal materials.

With the development first principles calculation of superconductivity based on electron-phonon coupling, subsequent calculations have validated Ashcroft's prediction: the strength of electron-phonon coupling of most metal calculations does not exceed 0.5, while the calculated strength of electron-phonon coupling of atomic metallic hydrogen is almost not less than 1.5, and the Debye temperature can reach 2000 K. For example, the metal hydrogen $I4_1/amd$ phase with the lowest stable pressure obtained from structural search is theoretically predicted to have an electroacoustic coupling strength of 1.81, The $\omega_{log}$ is 2068 K, so the superconducting temperature can reach as high as 315-356 K, making it a room temperature superconductor[58].

# 3. Structures and Superconductivity of Hydrides at High Pressure

Compared with metallic hydrogen, the threshold stable pressures of superconducting hydrides are much lower as the results of the chemical pressures, triggering the enthusiasm for hunting high-temperature superconductors in hydrides. Notably, theoretical simulations play a remarkable role in the discovery of superconducting hydrides via prediction of chemical components, crystal structures, stability and critical temperatures, which are rare in the previous research on superconductors. The section reviews the achievements of exploration of superconducting hydrides, including the typical structures and remarkably high-temperature superconductivity.

## 3.1 "Chemical pre-compression" theory

The concept of chemical pressure has a long history in condensed matter physics and was not only recognized in hydride theory research. For example, the first cuprate superconductor, lanthanum barium copper oxygen (LBCO), was discovered to have a $T_c$ of 30 K in 1986[65]. In January 1987, the enhanced $T_c$ of LBCO was discovered around 57 K at pressure ~ 1.68 GPa[66]. In March 1987, a strategy for chemical-pressure-enhanced $T_c$ was proposed for cuprate superconductor LBCO, suggesting to replace lanthanum with yttrium with a smaller radius to make the lattice of superconductor LBCO smaller, which can also achieve the effect of compressing the lattice and obtain the superconducting enhancement caused by pressed lattice[67]. Therefore, the superconducting critical temperature of yttrium barium copper oxygen (YBCO) under chemical pressure can reach 93 K, higher than the temperature range of liquid nitrogen. The idea of chemical pressure is generally based on atomic substitution in crystal structures to compress the lattice volume, resulting in lattice compression effects caused by external pressure, which is widely present in material design and discovery. However, the chemical pressure in hydrides is much more complex than that in other materials, because metallization of hydrogen is not only lattice compression, but also a polymerization of hydrogen molecular orbitals. The "chemical pre-compressing" elements in hydrides not only cause mechanical stress caused by replacements, but also electron transfer and orbital overlap between "pre-compressing" atoms and hydrogen, leading to decomposition and restructuring of orbital.

In order to achieve superconductivity of metallic hydrogen at lower pressures, Ashcroft proposed the "chemical precompression" theory 36 years after he prediction of superconductivity of metallic hydrogen[68]: when "chemical compressing" elements was be substituted into molecular hydrogen, the orbitals of "chemical compressing" elements would overlap with orbitals of molecular hydrogen, leading to the metallization of hydrogen in hydrides. Compared with metallic hydrogen driven by overlapping of orbitals of molecular hydrogen, it is expected that the pressure required for metallization of hydrogen in hydrides will be greatly reduced.

Concept of chemical precompression in 2004 fully demonstrated the foresight of theoretical physics, but there were a few limitations in the details suggested in the original literature[68], it is suggested in the original text to search for high-temperature superconductors in the hydrides of the fourth main group, which exhibit the highest hydrogen content at ambient pressure. However, in the further research, the search for hydrides in the fourth main group did not bring significant breakthroughs of remarkable high-temperature superconductivity[69-72]. As Ashcroft mentioned in the original literature: "*Unknown at present is the sequence of structures and states likely to be taken up by these hydrides upon systematic densification, including the possibility of layered arrangements and the intervention of intermediate phases of a partially ionic or charge-density-wave or even of semimetallic character*"[68], there were a few unresolved issues in the "chemical precompression" theory for hydrides. However, the accurate physical thinking of "chemical pre-compression" theory directly led to the experimental discovery of hydride superconductivity[44].

Compared to the widespread application of the "chemical pre-compression" theory in the theoretical design and experimental discovery of hydride superconductors over the past 20 years, the development of "chemical pre-compression" theory for hydrides has stagnated. At the current stage of the development of hydride superconductivity, further theoretical breakthroughs are needed to guide the theoretical design and experimental synthesis of hydride superconductors.

### 3.2 Hypervalent hydrides

Hypervalent structures refer to high-coordination structures formed by the main group elements with number of valent electron number exceeding eight. For example, phosphorus pentachloride $PCl_5$, sulfur hexafluoride $SF_6$, and xenon hexafluoride $XeF_6$ molecules are typical hypervalent structures[73]. A widely accepted theory is that additional electrons exceeding eight electrons occupied the non-bonded orbitals of the hypervalent structures. Because non-bond orbitals do not participate in bonding of structures, the energy of non-bond orbitals is similar to that of isolated atomic orbitals. Therefore, the chemical properties of hypervalent structures are generally determined by non-bonding orbitals. Due to the instability caused by the high energy of non-bonding orbitals, low energy of bonding orbitals is often required to balance the total enthalpy of hypervalent structures. Therefore, the fluorine with the highest electronegativity leads to many hypervalent fluorides, as shown in Fig. 5.

Hypervalent hydrides are a special type of hypervalent compounds. Although many chemical properties of hydrogen are similar to those of halogen elements, they all have one electron less to filled shells and tend to form diatomic molecules. But the electronegativity of hydrogen is much lower than that of fluorine. Because the central atom in most known hypervalent structures exhibits much lower electronegativity than the coordination atoms, hydrogen with relatively low electronegativity is difficult to form hypervalent structures. As early as 1975, the electronic structures of hypervalent molecules $SH_4$ and $SH_6$ were calculated[74] with high enthalpy. The calculated results suggest that both the hypervalent plane molecule $SH_4$ and the hypervalent octahedral molecule $SH_6$ have high formation enthalpies, because the electronegativity of hydrogen is smaller than that of sulfur. However, if the polarization effect of each sulfur-hydrogen bond is considered in the calculation, the decrease in enthalpy of the hypervalent octahedral molecule $SH_6$ (0.1954 Hartree) is three times higher than that of the normal molecule $SH_2$ (0.0457 Hartree). The decrease in enthalpy after considering polarity of bonds is not only caused by the number of polar bonds, but also the high symmetry of the hypervalent structure in terms of hydrogen bond polarity. Although hypervalent molecules generally have high-symmetry structures without polarity, when the hypervalent molecules were embedded into crystal structures, a decrease in the symmetry of hypervalent molecules will emerges, namely "steric effect" as shown in Fig. 5. From the crystal structures of $XeF_4$ and $XeF_6$, it can be inferred that the steric effect of the hypervalent structures is an important factor for high symmetry of the crystal structures.

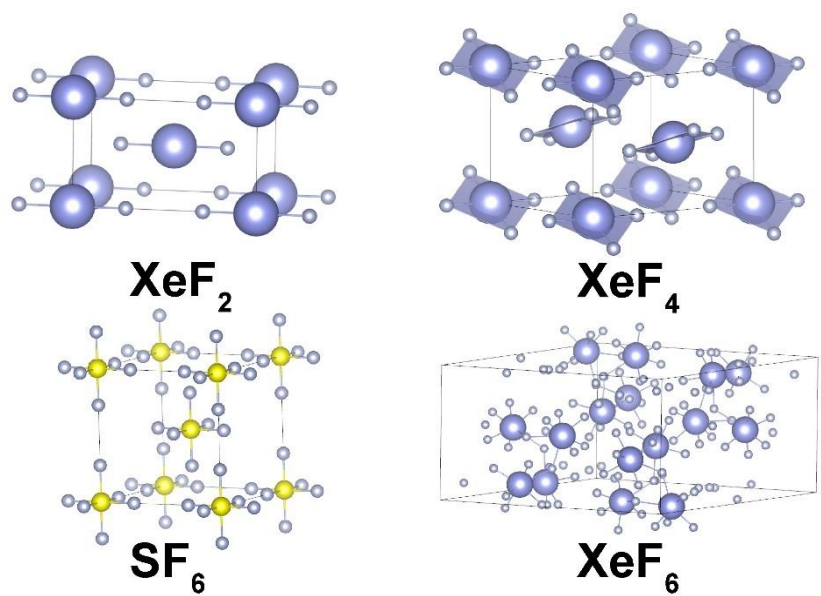

FIG 5| The crystal structures of typical hypervalent structures. $XeF_2$ is a linear molecule, and forms a linear molecular crystal after crystallization. $XeF_4$ is a square molecule, but it undergoes distortion due to steric effect after crystallization. $SF_6$ is a regular octahedral molecule, and forms a regular octahedral molecular crystal after crystallization. $XeF_6$ is a regular octahedral molecule, due to the stronger steric hindrance effect than $SF_6$ molecules, the shape of XeF6 regular octahedral molecules is severely distorted after crystallization.

The orbitals of octahedral molecule $SH_6$ occupied by valence electrons from sulfur and hydrogen are $a_{1g}$, $t_{1u}$, and $e_g$ [74]. The $a_{1g}$ and $t_{1u}$ orbitals are bonding orbitals that both sulfur and hydrogen contribute significantly, and the $e_g$ orbitals are mainly occupied by hydrogen valence electrons, namely non-bonding orbitals of hypervalent structure. Among these three orbitals, the orbital energy of $e_g$ is not only the highest (~ -0.3 Hartree), but also higher than the calculated energy of isolated hydrogen atomic orbitals (-0.5 Hartree). The energies of the other two bonding orbitals $a_{1g}$ and $t_{1u}$ occupied by the octahedral molecule $SH_6$ are -1.1 Hartree and -0.7 Hartree, respectively, which are lower than the calculated bonding orbital energies of the corresponding normal molecule $SH_2$ (the bonding orbitals $4a_1$, $2b_1$, $5a_1$, and $2b_2$ of SH2 have orbital energies of -1.0, -0.6, -0.5, and -0.4 Hartree, respectively). Therefore, it can be inferred that the high-energy non-bonding orbitals occupied in the hypervalent structures make it difficult to stabilize the super valence structure. However, the energy of the bonding orbitals in the hypervalent structures may be lower than that in the normal stoichiometric structures. Because the electronegativity of hydrogen is not high enough and the energy of the bonding orbitals is not low enough, the synthesis of hypervalent hydrides is difficult.

In 2012, synthesis of hypervalent hydrides was realized at high pressure. The synthesis and doping of the hypervalent hydride $K_2SiH_6$ were achieved under high temperature and pressure conditions using a large volume press[75], as shown in FIG 6. In the silicon-hydrogen system, the structural symmetry obtained by directly compressing silane is extremely low, and it may even dehydrogenate[71,76,77]. By utilizing potassium atoms to provide electrons and increase the distance between silicon-hydrogen units, the steric effect caused by polar bonds is weakened. The high-symmetry hypervalent hydrogen unit $SiH_6$ is successfully synthesized under high pressure and retained under ambient pressure. The non-bonding orbitals of the hypervalent hydrogen unit $SiH_6$, which do not participate in bonding, could exhibit the behaviors of atomic hydrogen. Moreover, the high energy of the non-bonding orbitals of hydrogen is balanced by the low energy of the bonding orbitals of the hydrogen hypervalent unit. The properties of hypervalent hydrogen seem to confirm Ashcroft's prediction of hydrogen-based superconductors in the "chemical pre-compression" theory. But unfortunately, the stoichiometric ratio of the hypervalent hydride $K_2SiH_6$ remains normal, so the non-bonding orbitals of hydrogen are completely occupied. Therefore, the hypervalent hydride $K_2SiH_6$ has no metallic properties and is a semiconductor, making it impossible to possess traditional superconductivity. The first hydrogen-based superconductor $H_3S$ is not only a hypervalent structure[12,78], but also similar to the coordination of $SH_6$ molecules predicted in 1975[74]. Moreover, the stoichiometric ratio is also abnormal, so the non-bonding orbitals of hydrogen were part-occupied and metallic. Therefore, hydride $H_3S$ exhibits high-temperature superconductivity as predicted in metallic hydrogen.

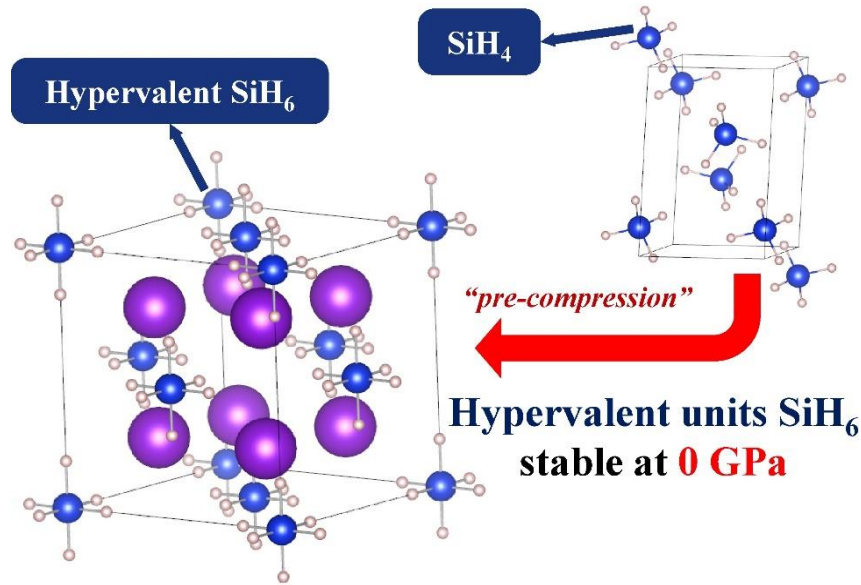

FIG 6 The first hypervalent hydride $K_2SiH_6$.

### 3.3 Typical hydride superconductors

Since the discovery of the high-temperature superconductivity of $H_3S$, the phase diagram of binary hydrides has almost been theoretically explored. Binary hydride superconductors can be roughly divided into two categories: covalent hydride superconductors and ionic polymeric hydride superconductors[79].

Covalent hydrides are widely discovered in nature, such as water and methane. However, these covalent hydrides are mainly insulators and cannot exhibits traditional superconductivity. In order to obtain metallization of atomic hydrogen in covalent hydrides, it is necessary to rely on the non-bonding orbitals of hypervalent hydrides discussed earlier. In 2014, the covalent hypervalent hydride $H_3S$ is theoretically predicted to have a superconducting critical temperature of 191-204 K at pressure ~ 200 GPa with the space group $Im\bar{3}m$ as shown in FIG 6[11,78]. Notably, the structure of hydride $H_3S$ is similar to the $SH_6$ molecule[74], where sulfur atom is bonded to six surrounding hydrogen atoms. In 2015, the $T_c$ of hydrogen sulfide samples was observed about 203 K at 155 GPa, which is very consistent with the theoretical prediction of the superconducting critical temperature of cubic hypervalent hydride $H_3S$[12]. In 2016, the high-temperature superconducting phase in hydrogen sulfide samples was determined from cubic hypervalent hydrides $H_3S$ through high-pressure in-situ electrical and synchrotron radiation XRD experiments[80]. In 2017, the in-situ spectroscopic studies on hydride $H_3S$[81] confirmed that the high-temperature superconductivity of $H_3S$ is dominated by electron-phonon coupling, which is consistent with theoretical predictions. Hypervalent hydride $H_3S$ is the first synthesized hydride high-temperature superconductor in high-pressure experimental, and also the first superconductor with a superconducting critical temperature above 200 K.

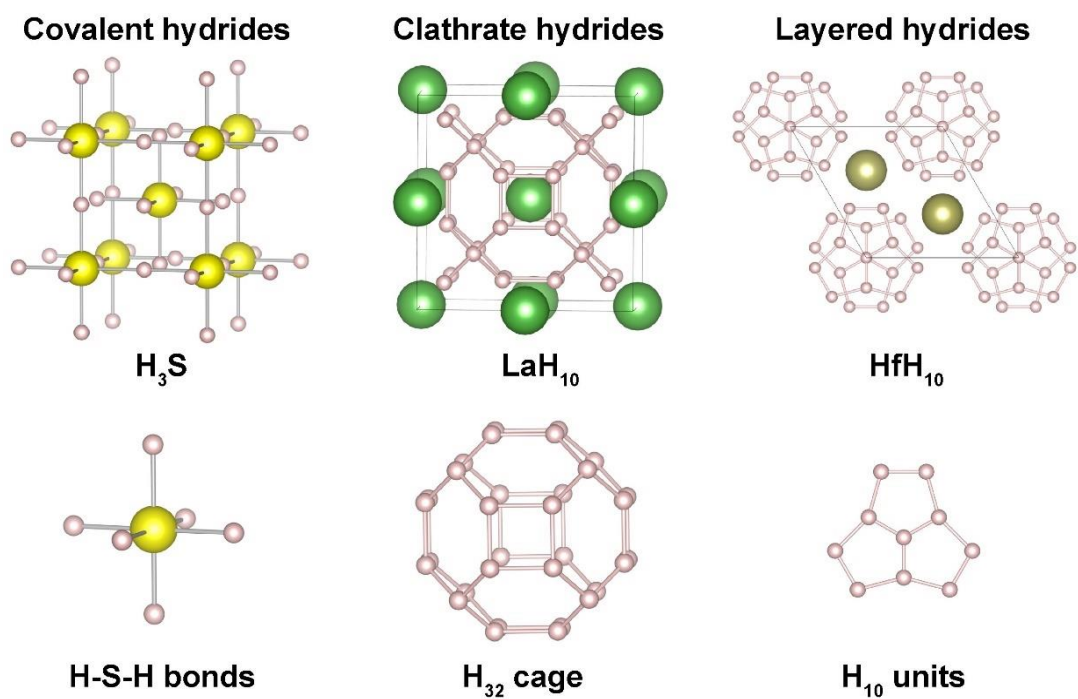

FIG 7 Typical crystal structures and hydrogen arrangements of binary hydrogen-based superconductors.

Although typical covalent binary hydride superconductors only include $H_3S$ and isomorphic $H_3Se$, and the $T_c$ of other covalent hydrides are generally not high[82], but we still could summarize rules from these covalent compounds. Firstly, when the radius of central atom of a hypervalent compound is too large, the steric effect of the hypervalent compound will be strong. Therefore, will the covalent bonds of the hypervalent units will be distorted in crystal caused by the steric effect. As shown in FIG 5, the small radius sulfur fluorine hypervalent unit $SF_6$ can crystallize into highly symmetric crystals, but the xenon fluorine XeF6 cannot form high symmetry crystals with large radius, and the $XeF_6$ hypervalent unit can only exist as gaseous molecules. Therefore, the radius of the central elements of hypervalent hydrides cannot be too high. However, as the radius of the element decreases, the electronegativity increases. If the electronegativity of the central elements is too high, it is impossible for the formation of hypervalent hydrides. Therefore, covalent binary hydride superconductors with hypervalent structures have requirements for the central elements, and only sulfur and selenium with suitable radius and electronegativity can meet the conditions for formation of cubic hypervalent lattice. However, the comparison of high-pressure experiments between silane and hypervalent hydride $K_2SiH_6$ indicates that the difficulty of formation of hypervalent units can be solved by introducing large radius elements into the crystal, as shown in FiG 6. Directly compressing silane can result in a structure with very low symmetry and even the dehydrogenation from silane. However, doping potassium ions to the lattice not only suppress the dehydrogenation, but also stabilizes the hypervalent unit $SiH_6$ at ambient pressure. Based on this approach, the selection range of central elements can be greatly expanded for hypervalent hydrides, inspiring the theoretical design rules and the discovery of the hydride $LaBeH_8$[18].

In addition to covalent hydrides, binary hydride high-temperature superconductors also have ionic hydride superconductors, their structural features are metal cations and polymeric hydrogen frameworks in crystal lattices. In 2012, the first ionic high-temperature superconductor $CaH_6$ was proposed with a body centered cubic structure and clathrate framework of polymeric hydrogen (space group $Im\bar{3}m$)[83]. In this coexisting structure of metal cations and framework of polymeric hydrogen, $H_2$ molecules undergo polymerization to form a connected hydrogen framework structure similar to polymerized nitrogen[59]. Because hydrogen has fewer valence electrons than nitrogen, the connected framework of polymerized hydrogen is metallic. Notably, the coordination of hydrogen in the hydride $CaH_6$ is the same as that of the $I4_1/amd$ phase of metallic hydrogen, both of which are 4-coordinated. However, the stable pressure of the hydride $CaH_6$ in the 4-coordinated polymeric hydrogen is much lower than that of the metallic hydrogen. Because metal Ca transfers electrons to hydrogen molecules, causing them to polymerize and form a covalent cage-like structure. This high-symmetry structure generates a Jahn Teller effect, which leads to phonon softening and promotes electron-phonon coupling, resulting in a superconducting critical temperature of up to 220-235 K for the hydride superconductor $CaH_6$. Then, the hydride $CaH_6$ was synthesized and observed a superconducting critical temperature of up to 215 K at 172 GPa[84]. When the reactants for the hydride $CaH_6$ is elemental Ca and hydrogen, Ca will be hydrogenated at low pressure to form impurities (such as $CaH_2$, $CaH_4$, and $Ca_2H_5$), which may hinder further hydrogenation for hydride $CaH_6$. In the synthesis of hydride $CaH_6$ in 2022, a mixture of calcium and $BH_3NH_3$ was used as the reactants, and first compressed the sample directly to 160-190 GPa. Then, the sample is heated by laser to release hydrogen from $BH_3NH_3$, and metal Ca is hydrogenated under high pressure, forming a clathrate hydride $CaH_6$.

The first clathrate hydride synthesized in the experiment was the hydride $LaH_{10}$, which has a face centered cubic structure (space group $Fm\overline{3}m$). And the theoretical prediction for hydride $LaH_{10}$ suggests a superconducting critical temperature ~ 286 K at 210 GPa[13]. At the same time, rare earth metal hydrogen rich systems were explored, proposing three clathrate hydrides that can widely exist in rare earth element hydrides with $H_{24}$ hydrogen cage, $H_{29}$ hydrogen cage, and $H_{32}$ hydrogen cage. Among them, $YH_{10}$ has a superconducting critical temperature ~ 303 K at 400 GPa[14]. In 2019, clathrate hydride $LaH_{10}$ was synthesized under conditions of 170 GPa and 1000 K with superconducting critical temperature ~ 260 K, which is close to theoretical predictions[15,85], proving that the theoretically predicted clathrate hydride is one of important families of hydride superconductors. In subsequent studies, many clathrate hydrides were synthesized experimentally, including $ThH_9$, $ThH_{10}$[86], $YH_9$, $YH_6$[87,88], $CeH_9$[89,90] and $CeH_{10}$[91].

In addition to clathrate hydrides, layered polymeric hydride superconductors are another typical ionic hydride family. Hydrogen molecules exhibits layered structures during compression, and both phase III and phase IV of solid hydrogen exhibit distinct layered feature. The structures of layered hydrides are the polymerization of layered hydrogen molecules into layered units, which exhibit the metallization of hydrogen. Therefore, in addition to the high-temperature superconductivity of hydrides, results of layered hydrides also help to understand the layered high-pressure phase of solid hydrogen. Layered polymeric hydrides were first discovered in the Sr-H system. The theoretical prediction of the superconducting transition temperature of layered hydrides $SrH_{10}$ at 300 GPa is 259 K[92,93], where hydrogen behaves similarly to phase III. In 2020, a new type of layered hydride $MH_{10}$ (M=Hf, Zr, Sc, Lu) was theoretically predicted with a pentagraphenelike $H_{10}$ layered unit[94], as shown in FIG 7. The superconducting critical temperature of the hydride $HfH_{10}$ with a pentagraphenelike structure at 250 GPa is around 213~234 K[94]. The prediction of the pentagraphenelike structure indicates that layered polymeric hydrides are the third family of hydride high-temperature superconductors besides covalent hydrides and clathrate hydrides. Pressure dependence of $T_c$s for the typical hydride superconductors are shown in FIG 8.

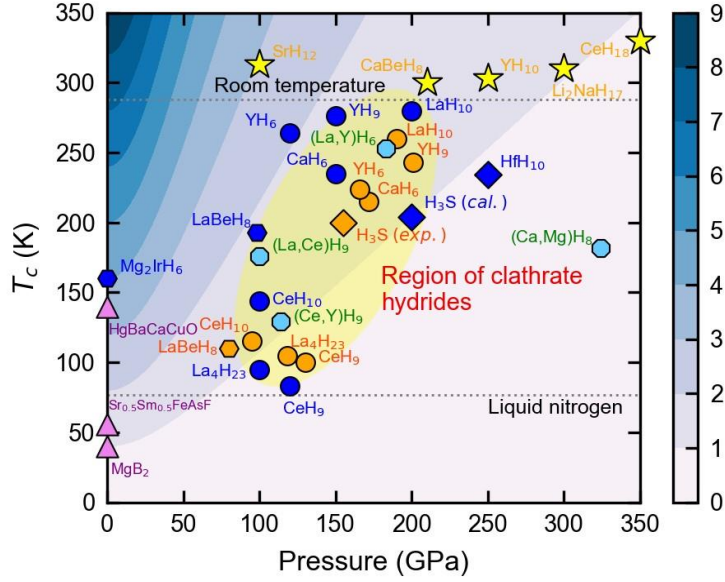

FIG 8. Pressure dependence of $T_c$s for typical superconductors. The orange and blue circles are $T_c$s of clathrate hydrides from experiments and theoretical predictions, respectively. The orange and blue hexagons are $T_c$s of ternary hydrides from experiments and theoretical predictions, respectively. The purple triangles are $T_c$s of well-known superconductors. The blue octagons are $T_c$s of ternary hydrides with solid solution structures from experiments. The yellow stars are potential room-temperature superconductors. The background is shaded according to the figure of merit $S = \dfrac{T}{\sqrt{P^2 + T_{MgB_2}^2}}$ used to evaluate the significance of a particular superconductor.

Currently, over 200 binary hydrogen-based superconductors have been predicted to be stable under high pressure and exhibit superconductivity[82]. However, there are still many deficiencies in the research of binary hydrides. Firstly, limited by the computational power of early hydride numerical simulations, early hydride structure searches rarely focused on non-integer stoichiometry, resulting in thermodynamic convex hull diagrams that might be insufficient. Insufficient searches for hydrides may lead to difference between results of theoretical simulation and experiments. Second, the calculations of non-hydrogen elements in different hydride systems did not consider the different properties of the elements, and different hydride systems often used the same set of calculation parameters. For example, how to consider the role of $f$ electrons for heavy rare earth elements in clathrate hydrides. Finally, the structure search of binary hydrides is based on the Born-Oppenheimer energy surface, and the influence of quantum and temperature effects on the structure search of hydrides has not been discussed yet.

Compared to binary hydride superconductors, ternary hydride superconductors have a more complex phase space, possibly forming superconductors with superior properties to binary hydrides. In particular, ternary hydrides have great promise to exhibit better superconductivity or stability under the synergistic effect of pre-compressed elements with different radii and electronegativity. But simulations for ternary hydrides requires more computational resources than that of binary hydrides. Therefore, research on ternary hydrides relies more on design theories and strategies, suggesting that the theory of "chemical precompression" needs further development to guide the theoretical design of hydrogen-based superconductors in ternary or even quaternary hydrides in the future. Although exploring ternary hydride superconductors remains significant challenges, many achievements have been made in designing high-temperature superconductors from ternary hydrides.

For example, in 2019, the strategy was proposed for obtaining high-temperature hydride superconductors by doping molecular hydrides with electrons. Lithium element was dopped to the molecular hydrides $MgH_{16}$, and the electrons transferred from lithium polymerized the hydrogen molecules into a clathrate structure. Finally, the hydrogen-based superconductor $Li_2MgH_{16}$ with the highest theoretical predicted superconducting transition temperature was designed, which was dynamically stable at 250 GPa and predicted a superconducting transition temperature of 473 K[95]. Moreover, solid solution is another route to stabilize superconducting ternary hydrides, such as $(La,Ce)H_9$[96,97], $(La,Y)H_9$[98], $(Ce,Y)H_9$[99] and $(Ca,Mg)H_8$[100] as shown in Fig. 8.

Recently, "hydrogen-based alloy backbone" theory was proposed for realization of hydrogen-driven superconductivity at moderate pressure by engineering binary X-H backbones, which are "chemical pre-compressed" by metal elements[18]. Based on this theory, a class of ternary hydrides $AXH_8$ (A=Sc, Ca, Y, Sr, La, Ba, X=Be, B, Al) with a fluorite-type alloy skeleton has been theoretically designed. Compared with other reported H-based superconductors, the predicted $AXH_8$ hydride exhibits dynamic stability at pressures far below its thermodynamic pressure, while maintaining the strong electron phonon coupling required for high-temperature superconductivity. Among them, the hydride LaBeH$_8$ dynamically stabilizes to 20 GPa, which is much lower than the pressure range required to stabilize typical hydrogen-based superconductors as shown in FIG 9. Future research on other ternary systems may use similar methods to determine high-temperature superconductors that are closer to environmental pressure. Therefore, design of hydrides with a hydrogen alloy backbone is an effective method to obtain hydrogen-based superconductors with low stable pressure. In hydrogen alloy backbone hydrides, small radius elements form alloy with hydrogen to introduce bonds of bonding states into the backbones, replacing some H-H bonds of anti-bonding states. The prediction of hydride LaBeH$_8$ has successfully inspired the its synthesis in experiments. In the thermodynamically stable pressure range of hydride LaBeH$_8$, this hydride has recently been successfully synthesized, confirming that the fluorite-type backbone hydride is the first synthesized ternary hydride template[23]. Notably, results of recent high-throughput calculations[19-22] fit the description of hydrogen alloy backbone hydrides, suggesting that it is possible to carry out high-throughput calculations using design strategy of hydride superconductors.

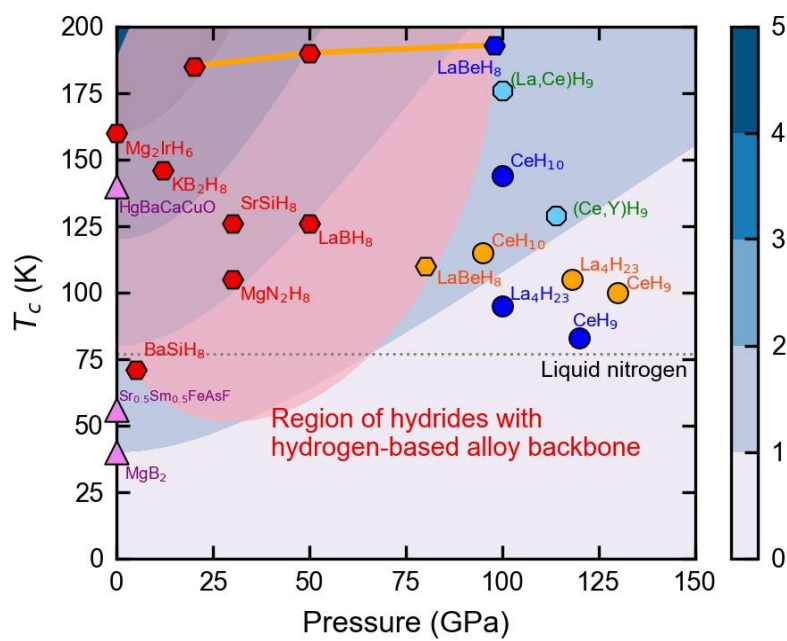

FIG. 9 Pressure dependence of $T_c$s for typical superconductors. The orange and blue circles are $T_c$s of clathrate hydrides from experiments and theoretical predictions, respectively. The orange and red hexagons are $T_c$s of ternary hydrides from experiments and theoretical

predictions, respectively. The blue hexagons is $T_c$s of hydride LaBeH$_8$ at thermodynamic pressure. The purple triangles are $T_c$s of well-known superconductors. The blue octagons are $T_c$s of ternary hydrides with solid solution structures from experiments. The background is shaded according to the figure of merit $S = \dfrac{T}{\sqrt{P^2 + T_{MgB_2}^2}}$ used to evaluate the significance of a particular superconductor.

## 4. Conclusions and Outlook

In the last 20 years, significant progress has been made in understanding of high-pressure structures and superconductivity of hydrogen and hydrides, thanks to the synergy between theory, numerical simulations and experiment. In this review, we mainly focused on what we learned from numerical simulations and how it guides the experiments of superconducting hydrides. An open question is how can we use chemical pressure to achieve superconductivity of metallic hydrogen at ambient pressure? It is generally believed that the pressure required by metallic hydrogen is higher than 500 GPa. As high-pressure experiments approach the stable pressure of metallic hydrogen and successful synthesis of polymeric hydrogen sublattice in clathrate hydrides, it can be predicted that the metallic hydrogen is the polymeric hydrogen molecular phase: the pressure greatly shortens the intermolecular distance, and hydrogen molecules undergo polymerization through multi-center bonding, similar to the observed polymeric nitrogen and carbon monoxide in experiments. Unlike olymerized nitrogen and polymerized carbon monoxide, polymerized hydrogen exhibits significant part-occupied multi-center bonds, suggesting a good metallicity. However, the stability of polymerized bonds is much lower than that of polymerized nitrogen and carbon monoxide, thus greatly increasing the difficulty of depressurization and retention of polymerized metal hydrogen. Although polymeric hydrides, such as clathrate hydrides, provide electron transfer to the polymeric hydrogen sublattice through "pre-compressed" atoms, greatly reducing the pressure at which polymeric hydrogen appears. However, electron transfer cannot effectively increase ability of hydrides to retain at low pressure, because electron transfer does not fundamentally change the bonding nature of polymeric hydrogen. As the results, the minimum pressure for dynamic stability of clathrate hydrides is not much lower than that for thermodynamic stability. Therefore, stabilizing the structure of polymerized hydrogen directly at ambient pressure is very challenging.

After a review of the chemical pressure of hydrides and numerical simulations of hypervalent hydrides, suggesting that a metallic hydrogen phase driven by chemical pressure does not exist in the elemental material, but exists in the non-bonding states of hypervalent units. For example, the $BeH_8$ unit in the "fluorite-type backbone" hydride $LaBeH_8$ and the $BeH_6$ unit in the hydride $Mg_2BeH_6$. There are two properties of hydrogen from hypervalent units: i) Low energy bonding orbitals bring excellent stability; ii) Part-occupied non-bonding orbitals cause the metallization of hydrogen. Due to these two properties, it is suggested to explore the hypervalent hydrides under chemical pressure in theoretical simulations and experiments for hydrogen-driven high-$T_c$ superconductivity at ambient pressure. Although realization of hydrogen-driven high-$T_c$ superconductivity still remains an open question and needs progress of both experiments and theoretical studies, the development of methods for synthesizing superconducting hydrides and the accumulation of related technologies have been steadily advancing. We believe that we are steadily approaching the theoretically proposed structures of hydrogen and hydrides with superconductivity.